\documentclass{svjour3}                     
\smartqed  
\usepackage{graphicx}
\begin{document}

\title{Dynamics of Charged Quantized Vortices 
}
\subtitle{Ion-vortex complexes}


\author{Makoto Tsubota         \and
        Hiroyuki Adachi 
}


\institute{M. Tsubota , H. Adachi\at
              Department of Physics, Osaka City University, Osaka 558-8585, JAPAN \\
              Tel.: +81-6-6605-3073\\
              Fax: +81-6-6605-2522\\
              \email{tsubota@sci.osaka-cu.ac.jp}           
}

\date{Received: date / Accepted: date}

\maketitle

\begin{abstract}
We study theoretically the dynamics of charged quantized vortices (CQVs). 
CQVs (ion-vortex complexes) have been used as an important probe in the field of quantum hydrodynamics. 
Recent experimental studies of quantum turbulence at very low temperatures utilize CQVs.
In this work we propose the equation of motion of CQVs based on the vortex filament model.
An analytical solution for a charged vortex ring shows how it expands under an electric field.
Numerical simulation reveals the characteristic behavior of CQVs under an electric field.
\end{abstract}

\section{Introduction}
\label{intro}
CQVs (ion-vortex complexes) have been used as an important probe in the field of quantum hydrodynamics since the pioneering work by Rayfield and Reif \cite{Rayfield}. 
Recent experimental studies on quantum turbulence at very low temperatures are performed by means of CQVs \cite{WG08}.
However, there has been very little theoretical information on the dynamics of CQVs \cite{BDV83,SD91}.
In order to understand the recent experiments, we need the information of the nonlinear dynamics of CQVs under an electric field.
In this work, first, we propose the equation of motion of CQVs on the vortex filament model. Then we discuss some typical analytical and numerical solutions.
\section{Equation of motion of CQVs}
\label{sec:1}
First, we remember how to derive the usual Schwarz's equation \cite{Donnelly}
\begin{equation}
\frac{d \,{\bf s}}{dt}={\bf v}_s+\alpha \,{\bf s}' \times ({\bf v}_n-{\bf v}_s)-\alpha' {\bf s}'\times [ {\bf s}' \times ({\bf v}_n-{\bf v}_s)]\, . \label{Schwarz}
\end{equation} 
Here ${\bf s}$ refers to a point on a vortex filament and ${\bf s}'$ is its derivative with respect to the coordinate along the vortex. The symbols ${\bf v}_s$ and ${\bf v}_n$ are the superfluid and the normal fluid velocity fields, and $\alpha$ and $\alpha'$ are the coefficients of the mutual friction.  The Magnus force per unit length is written as
\begin{equation}
{\bf f}_M=\rho_s \kappa {\bf s}' \times \left( \frac{d\,{\bf s}}{d\,t}-{\bf v}_s \right)
\end{equation}
with the superfluid density $\rho_s$ and the circulation quantum $\kappa$.
The drag force per unit length due to the mutual friction is 
 \begin{equation}
 {\bf f}_D=-\alpha \rho_s \kappa {\bf s}'\times [ {\bf s}' \times ({\bf v}_n-{\bf v}_s)]-\alpha'\rho_s \kappa {\bf s}' \times ({\bf v}_n-{\bf v}_s) \, .
 \end{equation}
Since the inertia of the vortex core is negligible, the equation of motion per unit length is ${\bf f}_M+{\bf f}_D=0$, which leads to Eq. (\ref{Schwarz}).  

Next, we consider the case that a part of a vortex is electrically charged. This describes an ion (the charge $e$ and the radius $R_{\rm ion}$) trapped by a vortex core, namely an ion-vortex complex.
An electric force ${\bf f}_E=e{\bf E}$ acts on the charged part.
Now the equation of motion is $2R_{\rm ion}({\bf f}_M+{\bf f}_D)+{\bf f}_E=0$, from which we obtain
\begin{equation}
\frac{d \,{\bf s}}{d\,t}={\bf v}_s+\frac{e}{2R_{\rm ion}\rho_s\kappa}{\bf s}'\times {\bf E}+\alpha \,{\bf s}' \times ({\bf v}_n-{\bf v}_s)-\alpha' {\bf s}'\times [ {\bf s}' \times ({\bf v}_n-{\bf v}_s)]\, . \label{Tsubota}
\end{equation}  
The second term of the r.h.s. is characteristic of the present case. This term is explicitly included in the vortex filament model for the first time. The physics of this term has been already discussed in some previous literature\cite{Rayfield,BDV83,SD91}, which was limited to simple symmetric cases.  In the rest of the paper we consider the case of zero temperature without mutual friction.

\section{Analytical solution of expansion of a charged vortex ring}
A key point of this issue is how the ion is localized along a hollow vortex core.  Early works speculated that the ion is in effect distributed uniformly along a core \cite{Rayfield}.  The observations of the ion mobility along the vortex core show, however, that the mobility increases rapidly with decreasing temperature \cite{OG76}. The physics of the behavior is not yet fixed. In this work we consider both cases of the uniformly distributed ion and the localized ion, not confining ourselves to either model. 

The model of the uniformly distributed ion enables us to obtain the analytical solution of how a charged ring expands under an electric field. 
Now we assume that initially a charged vortex ring (CVR) with a radius $R_0$ lie on a plane perpendicular to an electric field {\bf E} in absence of any flow ($v_n=v_s=0$). The charge $e$ is supposed to be uniformly distributed along the ring.
The equation of motion (\ref{Tsubota}) of a ring with a radius $R$ is reduced to
 \begin{equation}
 \frac{dR}{dt}=\frac{eE}{2\pi \rho_s \kappa R},
\end{equation}
whose solution is 
\begin{equation}
R(t)=\sqrt{\frac{eE}{\pi \rho_s \kappa}t+R_0^2} \, . \label{solution}
\end{equation}

\section{Numerical simulation of CVR dynamics}
We performed the numerical simulation of a CVR dynamics by the vortex filament model \cite{TAN00}, and found some behavior characteristic of the system. 
The simulation begins with a perfectly circular vortex ring.
The ion is assumed to be localized on a point on the vortex ring.
The ion is pulled by an electric field, exciting Kelvin-waves (KWs) with small amplitude. 
The KWs propagate along the vortex ring to keep approximately the circular ring shape.
As a result, the expansion of the ring with a localized ion is almost same as that with a delocalized ion.
     
Consider a CVR with radius $R=0.5$\,$\mu$m whose self-induced velocity is towards the $+z$ direction.
An applied electric field is $E{\hat z}$ with $E=100$\,V/cm, and the ion radius is $R_{ion}=15$\,nm.
These parameters follow the recent experiments by Walmsley and Golov \cite{WG09}. 
We made the numerical simulation by the vortex filament model. 
The motion of a charged part is described by Eq. ({\ref{Tsubota}), while other neutral parts move with the usual equation neglecting the second electric term.
The numerical space resolution is $\Delta \xi=3$\,nm, and the time resolution is $\Delta t=0.1$\,nsec.

Figure 1 shows the vortex configuration at $t=0.15$\,msec. Since the deformation along the $z$ axis is quite small, it is exaggerated by a factor of 500. Since the self-induced velocity of the ring is along the electric field, the charged part of the ion likes to expand through the second term of Eq. (\ref{Tsubota}). The resulting cusp excites KWs which propagate over the whole ring. However, these KWs are never amplified too much, so that the vortex ring propagates keeping the circular shape. This is consistent with the analysis based on the mode expansion \cite{SD91}.
 \begin{figure*}
  \includegraphics[width=1.\textwidth]{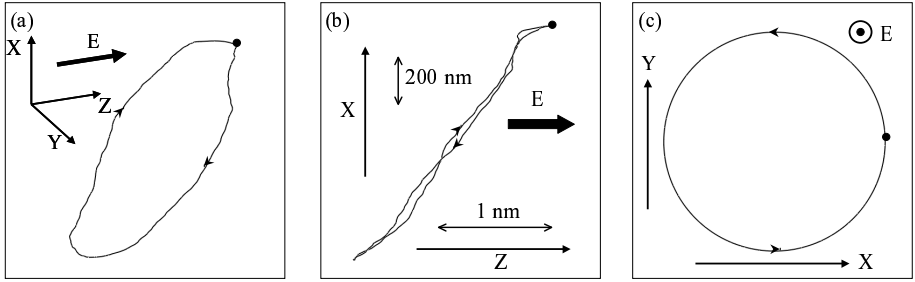}
\caption{Configuration of a vortex loop at $t=0.15$\,msec. These three figures depict the same vortex from different directions. The electric field ${\bf E}$ is applied along the $z$ axis. The small solid circle, located at rightmost extent of the vortex in each of the three frames,  shows the localized ion. The arrows on the vortex lines show the direction of circulation. The $z$ axis is exaggerated by a factor of 500. The ring radius in (c) is 0.5053$\,\mu$m. }
\label{Dynamics} 
\end{figure*}

Figure 2 shows the comparison of the radius growth between the solution of Eq. (\ref{solution}) for the delocalized-ion model and the numerical simulation assuming the localized-ion model.  The agreement is quite nice.  This is because the KWs starting from the ion soon cover the whole ring, as discussed in the next section. Although the observed time period is short, we should expect this agreement for longer time too. As the vortex ring becomes bigger, the propagation velocity is reduced.
 \begin{figure*}
  \includegraphics[width=0.6\textwidth]{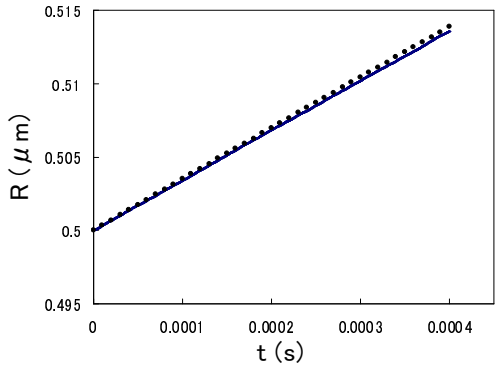}
\caption{Expansion of the ring radius. The solid line shows the numerical result for the localized-ion model, while the dashed line refers to the analytical solution of Eq. (\ref{solution}) for the delocalized-ion model. }
\label{Rt} 
\end{figure*}

\section{Discussions: the role of KWs}
It is well known that KWs play important roles in quantum turbulence at very low temperatures\cite{PLTP,TsubotaJPSJ}. So do they for the present problems.  The key question is how KWs transfer the energy from the work due to the electric field to the whole vortex ring to expand it.  

The early elementary process is clearly shown in Fig.3.  The electric field pulls the ion to excite a small cusp there. The resulting KWs have initially the wavelength comparable to the cusp scale, propagating apart from the ion.  The nonlinear interaction is known to work between different wavenumber modes of KWs and cause the KW cascade \cite{VTM03}. As the initial KWs propagate apart from the cusp, KWs with smaller wavenumbers are also excited through the nonlinear interaction to follow the initial KW.    Figure 4 shows how the KWs generated from the ion propagate and cover the whole ring. Through this process the energy injected from the ion becomes about uniformly distributed over the ring, so that the vortex keeps the ring shape. This is the reason why the radius growth of the solution of Eq. (\ref{solution}) for the delocalized-ion model agrees with that of the simulation using the localized-ion model (Fig.2).    
 \begin{figure*}
  \includegraphics[width=1.\textwidth]{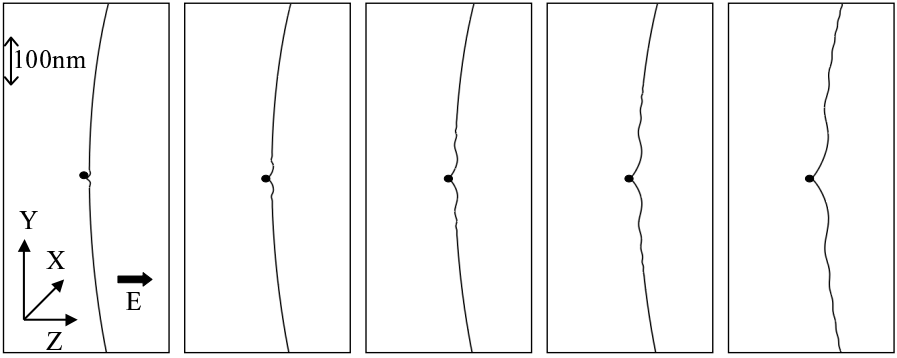}
\caption{Early elementary process of how the cusp grows from the ion and Kelvin-waves are excited. The ion is at the top of the cusp. The electric field is applied along the $z$ direction. The time is 0.7nsec, 1.7nsec, 4.0nsec, 7.0nsec and 10.0nsec from the left to the right. The $z$ axis is exaggerated by a factor of 500.}
\label{Dynamics} 
\end{figure*}

 \begin{figure*}
  \includegraphics[width=1.\textwidth]{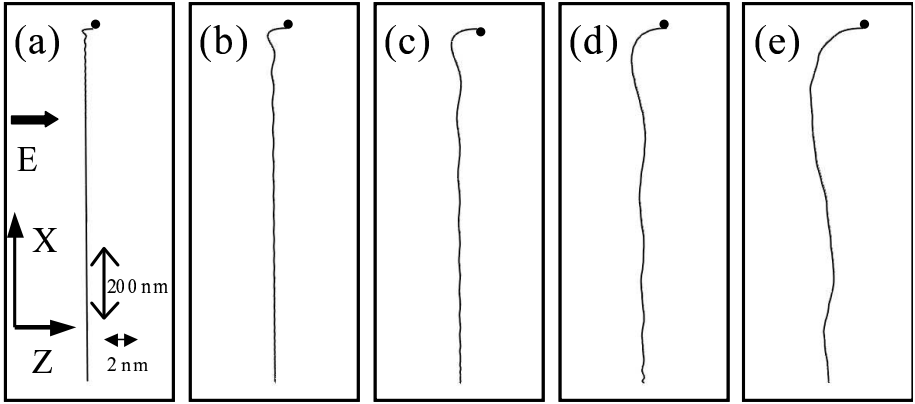}
\caption{Some late stage showing how the Kelvin-waves excited at the ion propagate and extend over the whole vortex ring. The solid circle refers to the ion. The time is 30nsec, 100nsec, 200nsec, 400nsec and 1000nsec from the left to the right. The $z$ axis is exaggerated by a factor of 500.}
\label{Dynamics} 
\end{figure*}
One mystery is why the KWs are not so much amplified in spite of the continuous excitation by the electric field. One reason would be that the energy of KWs is partly consumed in expanding the vortex ring. Another reason may come from the finite numerical space resolution. The simulation cannot describe the KWs with wavelengths shorter than the resolution and this works in effect as some dissipative mechanism, which occurs in the simulation of the KW cascade  \cite{VTM03}. 
  
In order to justify quantitatively the arguments in this section, we need the analysis of the spectra of KW during the dynamics, which is now in progress.  

\section{Conclusions}
We study theoretically and numerically the dynamics of charged vortex rings under an electric field.  First we introduce the characteristic term into the vortex filament model. Next the dynamics of a charged vortex ring is investigated by two models. One model assumes that the charge is uniformly distributed over the vortex because of the high mobility of the ion. Then we obtained the analytical solution of how the ring expands. The other model supposes that the ion is localized on the vortex.  The vortex filament simulation reveals how the electric field causes a cusp at the ion and the KWs are excited to extend over the vortex. 

These findings would be helpful for understanding the recent experiments of quantum turbulence using ions by Walmsley and Golov \cite{WG09}. It should be quite interesting and important to study the dynamics of multi-charged vortices and charged vortex tangles as studied by them. These developments are now in progress.

\begin{acknowledgements}
We would like to thank P. M. Walmsley and A. I. Golov for giving their preliminary experimental data and the fruitful discussions.
\end{acknowledgements}



\end{document}